\documentclass[journal]{IEEEtran}
\usepackage{cite}
\usepackage{graphicx}
\usepackage{amsmath}
\usepackage{amssymb}
\usepackage{algorithm}
\usepackage{fancyhdr}

\begin{document}

\footnote{Copyright (c) 2013 IEEE. Personal use of this material is permitted. However, permission to use this material for any other purposes must be obtained from the IEEE by sending a request to pubs-permissions@ieee.org.}

\title{Verification-Based Interval-Passing Algorithm for Compressed Sensing}


\author{Xiaofu~Wu and Zhen~Yang 

\thanks{This work was supported in part by the National Science Foundation of China under Grants 61032004, 61271335.
        The work of Yang was also supported by the National Science and Technology Major Project under Grant 2010ZX0 3003-003-02, and by the National Basic Research Program of China (973 Program) under Grant 2011CB302903.
        }
\thanks{Xiaofu~Wu and Zhen~Yang are with the Institute of Signal Processing and Transmission, Nanjing University of Posts and Telecommunications, Nanjing 210003, China (Emails:
        xfuwu@ieee.org, yangz@njupt.edu.cn)).}}


\maketitle

\begin{abstract}
We propose a verification-based Interval-Passing (IP) algorithm for iteratively reconstruction of nonnegative sparse signals using parity
check matrices of low-density parity check (LDPC) codes as measurement matrices. The proposed  algorithm can be considered as an improved IP algorithm by further incorporation of the mechanism of verification algorithm. It is proved that the proposed algorithm performs always better than either the IP algorithm or the verification algorithm.
Simulation results are also given to demonstrate the superior performance of the proposed algorithm.
\end{abstract}


\IEEEpeerreviewmaketitle

\section{Introduction}

\PARstart{C}{ompressed} Sensing (CS) problem in the noiseless setting considers the estimation
of an unknown and sparse signal vector $\mathbf{x} \in \mathbb{R}^N$ from
a vector of linear observations $\mathbf{y} \in \mathbb{R}^M$, i.e.,
\begin{equation}
  \label{eq:sys}
   \mathbf{y} = H \cdot \mathbf{x},
\end{equation}
where $H \in \mathbb{R}^{M\times N}$ is often referred as measurement matrix or $sensing$ $matrix$ and only a small number
(the sparsity index), $K<<N$, of elements of $\mathbf{x}$ are non-zero.
The set containing the positions of these elements is known as the support set, defined as $\mathcal{S}=\left\{i\in [1,N]:  x_i \neq 0\right\}$,
with cardinality $|\mathcal{S}|=K$. The sparsity is often defined as $k=\frac{K}{N}$.

The solution to this system of equations is known to be given by the vector that minimizes $\|\mathbf{x}_0\|_0$
 ($\ell_0$-norm) subject to $\mathbf{y} = H \cdot \mathbf{x}_0$, which is a non-convex
optimization problem. In \cite{DonohoCS}, it was  established that the vector $\mathbf{x}_1$ with
minimum $\ell_1$-norm subject to $\mathbf{y} = H \cdot \mathbf{x}_1$ coincides with $\mathbf{x}_0$
whenever the measurement matrix satisfies the well-known restricted isometry property (RIP) condition.

The connection between CS and Channel Coding (CC) has been explored extensively in recent years\cite{LdpcCS,ZhangCC_CS,ZhangCSIT,XuExpanderCS,DEVerfCS}. The sensing process in CS, i.e., $\mathbf{y} = H \cdot \mathbf{x}$, is very similar to encoding in CC if the $H$ is considered as a generator matrix while the reconstruction process in CS looks also similar to decoding in CC. Therefore, the sensing process in CS is often referred as $encoding$ and the recovery process as $decoding$.

In general, the sensing matrix in compressed sensing can be either $dense$ or $sparse$ according to the density of the nonzero entries in the sensing matrix.
As a special class of sparse matrices, the parity-check matrices of Low-Density Parity-Check (LDPC) codes have the favorite feature of a constant (average) number of non-zero entries in each row when the code length may increase without bound.  Due to the potential advantages on both $encoding$ and $decoding$, sparse sensing matrices have gained increasing interest in CS. With the bipartite-graph representation of any sparse measurement, various message-passing algorithms originally developed for decoding sparse-graph codes
have been introduced for reconstruction of sparse signals\cite{SarvCS,XuExpanderCS,DonohoMP,ZhangCSIT,BinGraphCS}.
The connection between LDPC codes and CS was addressed in detail in \cite{LdpcCS}.

Among various message-passing algorithms, the verification algorithm was first introduced for compressed sensing in \cite{SarvCS}. It is essentially identical to the earlier idea of verification decoding of packet-based LDPC codes \cite{LubyVB} as recognized by Zhang and Pfister \cite{ZhangCSIT}. This observation allowed a rigorous analysis of the verification decoding for compressed sensing via density evolution \cite{DEVerfCS}. The verification algorithm can be efficiently implemented with the complexity $O(N)$. In the literature, there are two categories of verification algorithms: node-based and message-based \cite{ZhangCSIT,DEVerfCS}. In general, the node-based algorithms have better performance and lower complexity.

For non-negative sparse signals, a new message-passing algorithm, referred as the Interval-Passing (IP) algorithm, was proposed in \cite{IPA1}, which can perform better than the verification algorithm for some LDPC-based sensing matrices \cite{IPAJounal}. The IP algorithm can be implemented efficiently with the complexity $O(N(\log(\frac{N}{K}))^2 \log(K))$.

In this paper, we focus on the two iterative message-passing reconstruction algorithms, i.e.,  the IP algorithm and the node-based verification algorithm. As the basic mechanism is somewhat different for the two message-passing algorithms, it remains open if one can find a message-passing algorithm taking advantages of both. Indeed, we propose a novel verification-based interval-passing algorithm for recovery of $nonnegative$ sparse signals, in which the mechanism of verification decoding can be concisely incorporated.

\section{Preliminaries}
\subsection{LDPC-Based Sensing Matrix and Sensing Graph}
In this paper, we consider a special class of sparse measurement matrices, the parity check matrices of binary LDPC codes.

It is well known that an LDPC code can be well defined by the null space of its sparse parity-check matrix. For any given sparse parity-check matrix, it can be efficiently represented by a bipartite graph. A bipartite graph has two sets of nodes representing the code symbols and the parity-check equations. These are called ``variable'' and ``check'' nodes, respectively.
An edge connects a variable node $n$ to a check node $m$ if and only if $n$ appears in the parity-check equation associated with $m$.

Similarly, a sparse sensing matrix has its own bipartite graph, which is referred as $sensing$ $graph$. Due to the close relation between CS and CC \cite{LdpcCS}, it is very helpful to describe iterative reconstruction algorithms over the sensing graph.

Let $G=(\mathcal{V}\bigcup \mathcal{C}, \mathcal{E})$ be the
bipartite graph of a sensing matrix $H=[h_{m,n}]$, where the set of variable nodes
$\mathcal{V}$ represents the signal components (or columns of $H$) and the set of
check nodes $\mathcal{C}$ represents the set of sensing constraints (or
rows of $H$) satisfied by the signal components.

Throughout this paper, we denote the set of variable nodes that participate in check $m$ by
$\mathcal{N}(m)=\{n: h_{m,n}\neq 0\}$. Similarly, we denote the set of
checks in which variable node $n$ participates as $\mathcal{M}(n)=\{m:
h_{m,n}\neq 0\}$.

\subsection{Interval-Passing Algorithm}
For nonnegative sparse signals, the IP algorithm iteratively computes the simple bounds of each signal component, i.e., both lower and upper bounds. With sufficient expansion guarantee for the sensing graph, it was proved in \cite{IPA1} that either the lower or the upper bound can converge to the same value for each variable node. Therefore, it can successfully estimate the sparse signal with a well-designed sensing graph.

The IP algorithm can be described in a rigorous message-passing form. It consists of two alternative update rules for messages along edges, one for check nodes and the other for variable nodes. In general, there are two bounding messages for any directed edge from $m$ to $n$, i.e., $L_{m\rightarrow n}$ for the lower bound and $U_{m\rightarrow n}$ for the upper bound.

Initially, $L_{n\rightarrow m} = 0, U_{n\rightarrow m}=y_m, \forall m\in [1,M], n\in [1,N]$. Then, two alternative update rules can be described as follows.

\textbf{Check node update}:
\begin{eqnarray*}
  \label{eq:cn}
     L_{m\rightarrow n} &=&\max \left\{0,y_m - \sum_{n'\in\mathcal{N}(m)\backslash n} U_{n'\rightarrow m}\right\} \\
     U_{m\rightarrow n} &=&y_m - \sum_{n'\in\mathcal{N}(m)\backslash n} L_{n'\rightarrow m}.
\end{eqnarray*}

\textbf{Variable node update}:
\begin{eqnarray*}
  \label{eq:cn}
     L_{n\rightarrow m} &=&\max_{m'\in\mathcal{M}(n)} L_{m'\rightarrow n} \triangleq L_n, \\
     U_{n\rightarrow m} &=&\min_{m'\in\mathcal{M}(n)} U_{m'\rightarrow n} \triangleq U_n,
\end{eqnarray*}

When it converges or the maximum number of iterations reaches, the IP algorithm finally outputs the decision vector $\hat{\mathbf{x}}=[\hat{x}_1,\hat{x}_2,\cdots, \hat{x}_N]$, where $\hat{x}_n = L_n, \forall n\in [1,N]$.

\subsection{Verification Algorithm}
For compressed sensing, the node-based verification algorithm works iteratively in a progressive way for recovering the signal components (variable nodes). Once a variable node is recovered, it is attached to the state of $verified$; otherwise, it is $unverified$. At the beginning of $decoding$, all the variable nodes are initialized as $unverified$. In each $decoding$ iteration, it employs three simple rules as follows.
\begin{itemize}
\item S1: If a measurement at ``check node'' $m$ is zero, all the  variable nodes neighbor to $m$ are verified as a zero value and labeled as $verified$.
\item S2: For a degree-1 measurement $y_m$, the variable node neighbor to $m$ is verified as the value of $y_m$ and labeled as $verified$.
\item S3: For a variable node $n$ subject to which there are $P\ge 2$ identical measurements ($y_m = \eta, \forall m \in \mathcal{M}_\eta (n)$  with $|\mathcal{M}_\eta (n)| = P$ for some $\eta\in \mathbb{R}$), all variable nodes neighboring a subset of these check nodes $\mathcal{M}_\eta (n)$ (not all of them) are verified with the value zero and labeled as $verified$.
\end{itemize}
Then, the verified variable node propagates its state to the other check nodes in its neighborhood so that these check
nodes can remove the contribution of this variable node from their respective measurement and try to infer, in the next
round, the values of the remaining variable nodes connected to them.

The above three rules in their rigorous form were reported in \cite{DEVerfCS}. The only difference here lies in the rule S3. In \cite{DEVerfCS}, it includes the mechanism that the
$unique$ variable node neighboring these check nodes is verified with the common value of the check nodes. This additional mechanism, however, is not required considering the rule S2. This rigourous form of S3 may not be fully recognized in \cite{ZhangCSIT,BinGraphCS} and the authors in \cite{BinGraphCS} rediscovered this mechanism to cope with cycles of length-4 in the sensing graph.

In what follows, we provide a precise implementation of the above three rules thanks to various formulations in \cite{DEVerfCS,BinGraphCS}. The node-based verification algorithm consists of two alternative update rules for node-based messages, one for check nodes and the other for variable nodes. The state of variable node $n$ is denoted as $s_n$. Then, $s_n=1$ is used to denote the variable node $n$ in a $verified$ state, and $s_n=0$ means $unverified$. For an efficient implementation of the rule S3, we introduce the state of coincidence \cite{BinGraphCS} with $s_n = -1$ for identifying the variable node $n$ when it accepts $P\ge 2$ identical measurements and $\xi_m = 1$ for signalling the check node $m, m\in \mathcal{M}_\eta (n) $. Let $\delta(x)=1$ if $x=0$ and zeros otherwise.

Initially, $\hat{x}_n=0, s_n =0, \forall n \in [1,N]$; $\xi_m=0, \forall m\in [1,M]$. Then, two alternative update rules for the node-based verification algorithm can be stated as follows.

\textbf{Check node update}:
\begin{eqnarray*}
  \label{eq:cn}
     \eta_{m} &=&y_m - \sum_{n\in\mathcal{N}(m)} \delta(s_n -1) \hat{x}_n, \\
     d_m &=& |\mathcal{N}(m)| - \sum_{n\in \mathcal{N}(m)} \delta(s_n -1), \\
  \mathcal{N}_0(m) &=& \{n: s_n =0, n\in \mathcal{N}(m)\}.
\end{eqnarray*}

\begin{itemize}
\item
IF ($\xi_m = 1$) and ($|\mathcal{N}_0(m)|\ge 1$)
\begin{eqnarray*}
    \forall n\in \mathcal{N}_0(m):  \hat{x}_n =0, s_n = 1; \\
    (\text{Signal release}):\xi_m = 0 .
\end{eqnarray*}
\end{itemize}

\textbf{Variable node update} (only for $n: s_n\neq 1$):
\begin{eqnarray*}
 \mathcal{M}_\eta (n)=\left\{m: m \in \mathcal{M}(n), \eta_{m}= \eta \right\}.
\end{eqnarray*}

\begin{itemize}
\item
IF ($\exists m\in \mathcal{M}(n), d_m = 1$)\footnotemark\footnotetext{If there are multiple such nodes, then choose one at random.}
\begin{eqnarray*}
  \label{eq:cn}
     \hat{x}_n &=& \eta_{m}, \\
     s_n &=& 1.
\end{eqnarray*}
\item
IF ($\exists m\in \mathcal{M}(n), \eta_{m} = 0$)
\begin{eqnarray*}
  \label{eq:cn}
     \hat{x}_n &=& 0, \\
     s_n &=& 1.
\end{eqnarray*}
\item
IF  ($\exists \eta \in \mathbb{R}, s.t. |\mathcal{M}_\eta (n)|\geq 2 $) and ($s_n = 0$)
\begin{eqnarray*}
  \label{eq:cn}
    s_n &=& -1, \\
    \xi_m &=& 1, \forall m\in \mathcal{M}_\eta (n).
\end{eqnarray*}

\end{itemize}

If all the variable nodes are verified, the algorithm outputs the recovered vector $\hat{\mathbf{x}}$.

\newtheorem{lem1}{Theorem}

\subsection{Remarks}
As shown, both the IP algorithm and the verification algorithm can be described in the form of message-passing decoding.
Although both algorithms require the expansion property of the sensing graph, their mechanisms in recovering the signal are somewhat different and simulations show their difference in recovery patterns.

Therefore, it is natural to ask if there is a message-passing algorithm in favor of the recovery mechanisms of both algorithms.

\section{Verification-Based Interval-Passing Algorithm}
In this section, we can answer the question in an affirmative manner. In what follows, we propose a verification-based interval-passing algorithm, which keeps all the messages appeared in the IP algorithm. Besides, it also introduces a new kind of message, $\eta_m$, attached to each check node $m$, which do the same role as in the verification algorithm.

Initially, $s_n = 0, \hat{x}_n =0, \forall n\in [1,N]$ and $\xi_m=0, \forall m\in [1,M]$. The messages along any edge from variable node $n$ to check node $m$ are initialized as $L_{n\rightarrow m} = 0$, $U_{n\rightarrow m} = y_m$. Then, the new algorithm iteratively does the following update operations in an alternative manner.

\textbf{Check node update}:
\begin{eqnarray*}
  \label{eq:cn}
     L_{m\rightarrow n} &=&\max \left\{0,y_m - \sum_{n'\in\mathcal{N}(m)\backslash n} U_{n'\rightarrow m}\right\}, \\
     U_{m\rightarrow n} &=&y_m - \sum_{n'\in\mathcal{N}(m)\backslash n} L_{n'\rightarrow m}, \\
     \eta_{m} &=&y_m - \sum_{n\in\mathcal{N}(m)} \delta(s_{n}-1) \hat{x}_n, \\
  \mathcal{N}_0(m) &=& \{n: s_n =0, n\in \mathcal{N}(m)\}.
\end{eqnarray*}

\begin{itemize}
\item
IF ($\xi_m = 1$) and ($|\mathcal{N}_0(m)|\ge 1$)
\begin{eqnarray*}
    \forall n\in \mathcal{N}_0(m): L_{m\rightarrow n} = U_{m\rightarrow n} = 0, \\
    \hat{x}_n = 0, s_n = 1;\\
    (\text{Signal release}):\xi_m = 0.
\end{eqnarray*}
\end{itemize}

\textbf{Variable node update} (only for $n: s_n\neq 1$):
\begin{eqnarray*}
  \label{eq:cn}
     L_{n\rightarrow m} &=&\max_{m'\in\mathcal{M}(n)} L_{m'\rightarrow n} \triangleq L_n, \\
     U_{n\rightarrow m} &=&\min_{m'\in\mathcal{M}(n)} U_{m'\rightarrow n} \triangleq U_n, \\
    \mathcal{M}_\eta (n)&=&\left\{m: m \in \mathcal{M}(n), \eta_{m}= \eta \right\}.
\end{eqnarray*}
\begin{enumerate}

\item
IF  ($\exists \eta \in \mathbb{R}, s.t. |\mathcal{M}_\eta (n)|\geq 2 $) and ($s_n = 0$)
\begin{eqnarray*}
  \label{eq:cn}
    s_n &=& -1, \\
    \xi_m &=& 1, \forall m\in \mathcal{M}_\eta (n).
\end{eqnarray*}

\item
IF ($L_n = U_n$)
\begin{eqnarray*}
   \hat{x}_n &=& L_n, \\
   s_n &=& 1.
\end{eqnarray*}

\end{enumerate}

When it converges or the maximum number of iterations reaches, the algorithm finally outputs the decision vector $\hat{\mathbf{x}}=[\hat{x}_1,\hat{x}_2,\cdots, \hat{x}_N]$, where $\hat{x}_n = L_n, \forall n\in [1,N]$.

\begin{lem1}
For recovery of nonnegative sparse signals with continuously-distributed nonzero entries, the new algorithm performs always better than either the IP algorithm or the verification algorithm.
\end{lem1}
\begin{proof}
Given a nonnegative sparse signal $\mathbf{x}$ and any measurement vector $\mathbf{y}=H \mathbf{x}$.  Let $\mathcal{V}^l=\{n: s_n=1\}$ denote the set of verified nodes at the $l$th decoding iteration. Hence, it is enough to prove that $\mathcal{V}_{new}^l \supseteq \mathcal{V}_{VB}^l$ and  $\mathcal{V}_{new}^l \supseteq \mathcal{V}_{IP}^l$, where the subscripts $new$, $VB$ and $IP$ mean the new algorithm, the verification algorithm and the IP algorithm, respectively.

Firstly, the probability of false verification has been demonstrated to be zero for the verification algorithm if the nonzero entries of the source signal are continuously distributed \cite{DEVerfCS}. For the IP algorithm, it always output the correct decision $\hat{x}_n=x_n$ if it converges at some variable node $n$ since $L_n \leq x_n \leq U_n$ always holds in the decoding process. Hence, it follows that the probability of false verification is zero for the new algorithm. Now, it is clear that $\mathcal{V}_{new}^l \supseteq \mathcal{V}_{IP}^l$ as no error propagation occurs.

Secondly, the two rules (S1/S3) employed by the verification algorithm still hold for the new algorithm. Hence, it is enough to show that the rule S2 still holds. Indeed, whenever $d_m = 1$ for any check node $m$, it means that there are $|\mathcal{N}(m)| -1$ $verified$ variable nodes neighbouring to $m$. Let us denote $n^*$ as the remaining unverified node (only one) neighbouring to $m$. Hence, $\forall n\in \mathcal{N}(m)/n^*,  L_n =U_n$ as it is verified. With the update operations at check node $m$, $L_{n^*}=U_{n^*}$ holds surely. It follows that $\mathcal{V}_{new}^l \supseteq \mathcal{V}_{VB}^l$.
\end{proof}

{\ }

Comparing the proposed algorithm with the IP algorithm, it is straightforward to show that the complexity of the proposed algorithm scales the same as that of the IP algorithm. For the quasi-cyclic (QC) LDPC-based sensing matrices, it may be implemented in a hardware-friendly manner as shown in decoding of QC-LDPC codes\cite{DecoderLDPC}.

\section{Simulation Results}
In this section, we provide the simulation results for three iterative reconstruction algorithms. The parity-check matrices of binary LDPC codes are adopted as sparse measurement matrices. The number of maximum $decoding$ iterations is set to 50 for all simulations.  In all simulations, sparse signals are generated as follows. First, the support set of cardinality $K$ is randomly generated. Then, each nonzero signal element is drawn according to a standard Gaussian distribution. If the generated element is negative, it is simply inverted for ensuring a nonnegative sparse signal model with continuously-distributed nonzero entries. For getting a stable point in all figures (except that the probability of correct reconstruction is approaching 1), at least 100 reconstruction fails are counted.

\begin{figure}[htb] 
   \centering
   \includegraphics[width=0.5\textwidth]{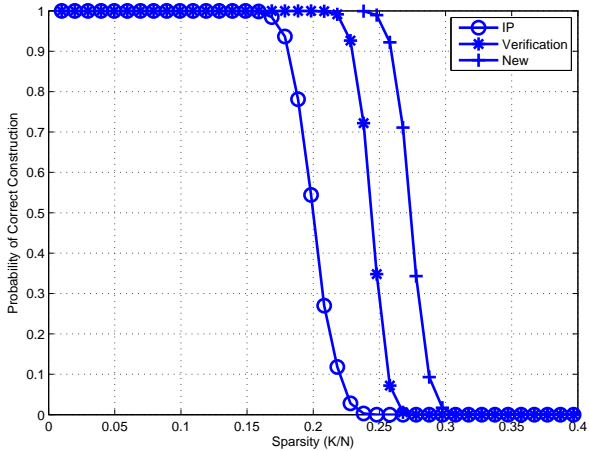} 
   \caption{Reconstruction of $k$-sparse signals with the MN-LDPC measurement matrix of length 504 ($N=504$, $M=252$).}
   \label{fig:1}
\end{figure}

Firstly, the binary MacKay-Neal LDPC matrix \cite{MacKay-Neal} of size $M\times N = 252 \times 504$ is adopted.
The probability of correct reconstruction is plotted against the sparsity of the source signals. As shown in Fig. \ref{fig:1}, the new algorithm performs always better than either the IP algorithm or the verification algorithm. It should be noted that this sensing matrix is constructed without length-4 cycles.

\begin{figure}[htb]
   \centering
   \includegraphics[width=0.5\textwidth]{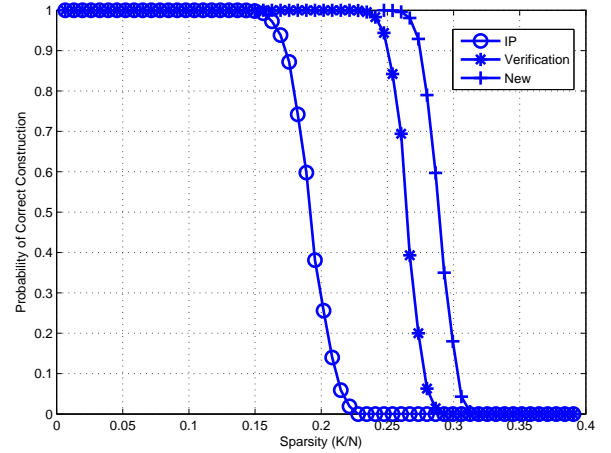}
   \caption{Reconstruction of $k$-sparse signals with the QC-LDPC measurement matrix of length 768 ($N=768$, $M=384$).}
   \label{fig:2}
\end{figure}

Secondly, we adopt the QC-LDPC code from the IEEE 802.16e standard, referred as the WiMax code. The parity-check matrix is of size $M\times N =384  \times 768$ and it has cycles of length-4. The same phenomena has been observed in Fig. \ref{fig:2}, which is clearly different with the results reported in \cite{IPAJounal}, where the verification algorithm performs poorly due to lack of mechanism to cope with cycles of length-4.

\section{Conclusion}

We have proposed a new message-passing reconstruction algorithm for nonnegative sparse signals.
This new message-passing algorithm, as a combination of  both the IP algorithm and the verification algorithm, has been shown to perform better than either the IP algorithm or the verification algorithm for LDPC-based sensing matrices. The complexity of the new algorithm scales the same as that of the IP algorithm.

\appendix[Response to one of the reviewers for the implementation of the S3 rule]
One of the reviewers has the following comments.

``The verification algorithm discussed is only equivalent to the S3 rule if the edge weight of the sensing graph comes from a continuous distribution (or its sampled version with infinitely many elements). Otherwise, it is very easy to come up with examples that the algorithm proposed there does not perform like the S3 rule. Indeed, it is very easy to show that in order to capture the S3 rule, one needs a memory state with 2 levels (variable nodes knowing about other variable nodes and check nodes knowing about other check nodes). The only way that you can implement the S3 rule with one level of memory (message passing) is to have continuous distribution on the edges of the sensing graph."

\begin{figure}[htb]
   \centering
   \includegraphics[width=0.5\textwidth]{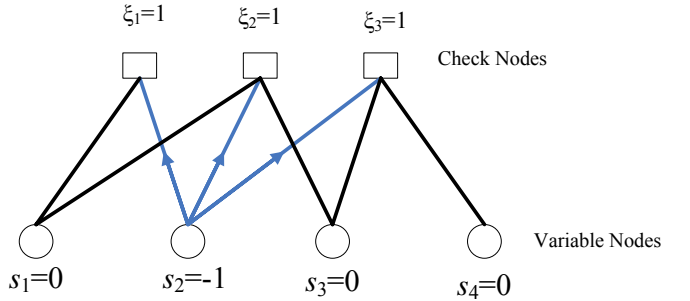}
   \caption{Illustration of the implementation of the S3 rule in verification algorithm.}
   \label{fig:3}
\end{figure}

In this paper, we always assume a binary edge weight for LDPC-based sensing graph. Consider the case of Fig. \ref{fig:3} with LDPC-based sensing graph. For the variable node $n=2$, there are $P=3$ identical measurements $\eta_1=\eta_2=\eta_3=\eta$. It is clear that the verification algorithm proposed in Section-II can implement the S3 rule precisely if the variable node $n=2$ is firstly visited. As shown, $\mathcal{M}_\eta(n)=\{1,2,3\}$.  Then, according to the S3 rule, the variables $n=1,3,4$ are verified with the value zero. This can be precisely implemented. However, it requires that the variable nodes are visited according to a descending order of the number of identical measurements inherited to each variable node. In practice, this can be omitted as the occurrence of $P>2$ identical measurements is rare.


\end{document}